\documentclass[preprint,12pt]{elsarticle}
\usepackage{amssymb}
\usepackage{epsfig}
\usepackage{graphicx}
\usepackage{color}

\begin{document}
\begin{frontmatter}

\title{Clustering large number of extragalactic spectra of galaxies and quasars through canopies}
\author[label1]{Tuli De}
\author[label2]{Didier Fraix Burnet}
\author[label3]{Asis Kumar Chattopadhyay}
\address[label1]{Department of Mathematics (M.Tech and B.Tech
unit), Heritage Institute of Technology,Kolkata,Chowbaga Road,
Anandapur, Kolkata 700 107,India; tuli\_stat5@yahoo.co.in}
\address[label2]{UJF-Grenoble 1 /CNRS-INSU,
Institut de Plan$\acute{e}$tologie et d'Astrophysique de Grenoble
(IPAG), UMR 5274, Grenoble, F-38041, France;
didier.fraix-burnet@obs.ujf-grenoble.fr}
\address[label3]{Department of Statistics, Calcutta University, Kolkata; 35, Ballygunge Circular Road, Kolkata - 700019, India; akcstat@caluniv.ac.in}

\begin{abstract}
Cluster analysis is the distribution of objects into different
groups or more precisely the partitioning of a data set into
subsets (clusters) so that the data in subsets share some common
trait according to some distance measure.Unlike classification, in
clustering one has to first decide the optimum number of clusters
and then assign the objects into different clusters. Solution of
such problems for a large number of high dimensional data points
is quite complicated and most of the existing algorithms will not
perform properly. In the present work a new clustering technique
applicable to large data set has been used to cluster the spectra
of 702248 galaxies and quasars having 1540 points in wavelength
range imposed by the instrument. The proposed technique has
successfully discovered five clusters from this 702248$X$1540 data
matrix.
\end{abstract}

\end{frontmatter}
\section{ Introduction}
Clustering is a technique used to place data elements into related
groups without advance knowledge of the group definitions.
Clustering algorithms are attractive for the task of
identification in coherent groups for existing data sets. It is a
common problem in several streams of science although the purpose
and implication may vary. Depending on the assumption and nature
of the data several techniques have been developed by scientists
for clustering in order to make proper analysis of data. Such
datasets are generally multivariate in nature. The common problem
is to find a suitable representation of the multivariate data.
Different clustering algorithms like k-means (MacQueen (1967)
\cite{MacQueen}), hierarchical methods like agglomerative and
divisive, model based clustering under distributional
assumptions,Bayesian methods etc. have been widely used by several
statisticians in order to identify the internal grouping
structures of objects. Due to unusual  nature of the data sets,
standard techniques often fail to identify the proper clusters.
For example, for heavy tailed distributions the distance based
classifiers with L2 norm can suffer from excessive volatility.
Hall (2009)\cite{Hall} suggested a new classifier by mixing L1 and
L2 norms where means and medians of marginal distributions take
different values. However in very high dimensional settings
conventional spatial median will create problems. For this
componentwise medians were
used to construct robust classifier.\\
Dimension reduction techniques can be used as an initial step in
statistical modelling and clustering.  Some dimension reduction
techniques like Principal Component Analysis (PCA) and Independent
Component Analysis (Common (1994) \cite{comon}) have been used for
clustering and identification of proper variables for the purpose
of grouping. Xia  (2008)\cite{Xia} suggested one method of
specification that involves separating of the linear components
from the non linear components,
leading to further dimension reduction in the unknown link function and, thus, better estimation and easier interpretation of the model.\\
 Depending on the physical situation, several limitations and specialities in the data create the need for the development of new techniques
 appropriate for those situations. In biostatistics, the situation where the number of variables is much larger than the number of observations
  is quite common and some new methods have already been developed.In such cases some initial reduction in the dimensionality is desirable before
  applying any PCA type method.  Johnstone et al.(2009)\cite{Johnston1}  suggested a simple algorithm for selecting a subset of coordinates with
  large sample variances and proved that if PCA is done on the selected subset then consistency is  recovered even if p is much grater than n. Several works
  (Johnstone et al. (2004)\cite{Johnston} , Zou et. al(2006)\cite{Zou}) have been reported on Sparse Principal Component Analysis. Jolliffe et al(2003 \cite{Jolliffe})
  suggested a modified PCA based on the LASSO.When both the number of variables and number of observations are quite large,
  most of the standard techniques will fail. In particular for non Gaussian situation those problems are quite apparent.\\
 Under astrostatistics, applications of dimension reduction and clustering techniques are quite common but in such applications traditional techniques
  often fail to explain the situation properly due to several unusual features like non Gaussianity, presence of outliers etc.
  Chattopadhyay et. al ( 2009) \cite{Chattopadhyay3},2012b \cite{Chattopadhyay2},2013 \cite{Chattopadhyay}) and Chattopadhyay et. al
  (2012a)\cite{Chattopadhyay1}
  considered some of the above problems.\\
Automatic sky surveys have begun to produce huge databases and
forthcoming telescopes will definitively lead the astrophysical
science into the era of big data requiring a renewal of
traditional approaches to tackle data. In Statistics, very often
we face empirical and large datasets to analyze. However, when
clustering algorithm are applied to large datasets,minimal
requirements of domain knowledge is necessary to determine the
input parameters.Further such algorithms should be able to
discover clusters with arbitrary shape and determine  the optimum
number of homogeneous classes automatically.Popular clustering
techniques such as the K-Means Clustering and Expectation
Maximization (EM) Clustering, fail to give solution to the
combination of these requirements. Also in recent days many
important
problems involve handling of large data sets where the standard clustering techniques fail or become computationally expensive.\\
Thus keeping in view the above considerations some new approaches has been developed which may be viewed as unsupervised clustering techniques or
Data Mining Approaches. These approaches mainly deal with large datasets. Data sets can be large in three ways. Firstly,there can be a large number of
elements in the data set,secondly each element can have many features,and finally there can be many clusters to discover.\\
 The objective of the present work is to perform a cluster analysis of a large sample of extragalactic spectra in order to identify important families
of galaxies and quasars.A first and similar work has been
performed in Sanchez-Almedia et al (2010) \cite{Sanchez Almeida}
using a k-means analysis. However, their clustering is not very
robust in the sense they could not conclude objectively on a
particular partitioning. In order to make the works comparable,
spectra have been
 selected from the same database, and used the same spectral bands to make the number of parameters tractable.\\
 In this paper a new clustering method is discussed that is efficient in handling data sets which may be considered as large in all three above
 mentioned
  respects.This method was primarily introduced by McCallum (2000)\cite{McCallum}.In this work, the algorithm is properly designed in order to apply it to the
  spectra data.The concept related to choice of the "cheap distance measure" is considered from a new angle.\\

  \section{The Canopy Clustering Technique}

 The technique mainly divides the data set into overlapping subsets termed as "canopies" based on a "cheap distance measure"
 (to be explained in later sections). Then in the second stage of the method, clustering is performed by measuring exact distances
 only between points which belong to a common canopy. Using canopies large data set clustering techniques can be performed with great
  efficiency even after allowing a small proportion of overlaps (common points between canopies) which were formerly impossible. Also,
  under reasonable assumptions, appropriate selection of cheap distance metric reduces computational cost without any loss in clustering accuracy.
   Canopies can be applied to many domains and used with a variety of clustering approaches such as Greedy Agglomerative Clustering, K-Means and
   Expectation-Maximization.

 \subsection{Creating the canopies}
 The key idea of the canopy algorithm is that one can greatly reduce the number of distance computations required for clustering by first cheaply
  partitioning the data into overlapping subsets i.e., the canopies.
 \newline A canopy is simply a subset of the elements  (the data points) that, according to the approximate similarity measure, are within some
 distance threshold from a central point. So, an element may appear in more than one canopy and every element must appear in at least one canopy.

 \subsection{Cheap distance Metric}

 The idea of cheap distance measure comes from the point of view that when the data set is too large in terms of both number of observations and
 dimensions, in that situation it is very difficult to compute an exact distance metric. In such cases, the idea is to select a distance measure
 which will be computationally simple and cheap. This distance metric may be selected in many ways. In situations where there are many variables or
  parameters in the data set, pairwise distance is calculated only for variable(s) which is (are) of most importance
  ( or count(s) for the maximum variability). Another possible way may be to reduce the dimension of the data by some dimension reduction
  technique and select the distance metric based on the reduced data set. After selecting the variable(s) which consists of the cheap distance metric,
   an observation point is chosen at random and the from that point distance (Euclidean/Manhattan) of all other points are calculated.

 \subsection{Selection of Distance thresholds}

 Given the above distance metric, the canopies are created by fixing two distance thresholds, may be called $T_{1}$ and $T_{2}$ such that
  $T_{1} > T_{2} $.Then a data point is selected at random and its distance is measured approximately to all other points. All the points that
  are within the distance threshold $T_{1}$ are put into a canopy and all the points which are within distance threshold $T_{2}$ are removed.
  These steps are repeated until the list is empty and data gets subdivided into overlapping canopies.

  \vskip10pt Since the selection of distance threshold is arbitrary (user defined), different cross validation techniques such as Time complexity,
  Calculation of Precision and Recall may be adopted to fix the proper choices of $T_{1}$ and $T_{2}$. Computational evaluation shows that the more
  accurate the selection of distance thresholds, the less is the overlaps between canopies and the better is the clustering accuracy.

  \vskip10pt  In the present work, the cross validation technique based on precision and recall is used.
  Precision is the fraction of correct predictions among all points predicted to fall in the same cluster.
 Recall is the fraction of correct predictions among all points that truely fall in the same cluster.
  \vskip10pt Some trial pair of values of  thresholds $T_{1}$ and $T_{2}$ are selected arbitrarily and canopies are formed based on the distance
  thresholds at each trial. Then for each individual selection, a discriminant analysis is performed assuming each canopy consists of a group.
  Then precisions and recalls are calculated for each groups or canopies. The total precision and recall is calculated by summing up the individual
  values.
  Then the pair of values of  $T_{1}$ and $T_{2}$ is chosen as the one for which precision and recall values are close to each other and simultaneously the proportion of correct
  observations is also highest.

\subsection{2nd Stage: subdivision of canopies}
 Once the canopies are formed in the first stage of the clustering, the major canopies are identified (i.e., the canopies containing larger number of
 observations). Then standard clustering approaches such as K-Means,Greedy Agglomerative or Expectation-Maximization techniques may be applied
 for further subdivision of the larger canopies. In these case expensive (actual) distance comparisons are performed within elements of a canopy.
\newline There are many methods for selecting the estimates of the cluster centroids named as prototypes.

\vskip10pt One of the approaches of selecting prototypes for further clustering, is to restrict the selection within points of a particular canopy.
As a result different prototypes are selected from different canopies.
\vskip10pt Another way may be to select prototypes based on the entire data points instead of restricting the selection of prototypes within a
particular canopies like the previous method.

 \vskip10pt In this paper, the first method has been adopted for selecting prototypes and K-Means clustering technique has been used for further
 subdivision of major groups or canopies.

\section{Computational Complexity of The Canopy Method:}

Since the canopy technique uses the cheap distance metric at the first step, it can reduce the computational complexity of a clustering algorithm
as a whole. Assuming that there are n data points,c canopies are formed based on the distance thresholds ,after formation of canopies there are a total number of N data points (including overlap or loss or both) and each data point falls into say m canopies on an average and the canopies are of equal size ,then there will be approximately mN/c data points per canopy. Hence, the total number of distance comparisons at this stage will be almost O(c(mN/c)$^{2}$) = O($m^{2}$$ N^{2}$/c).

In the next step, clustering is performed within each canopy  using the K-Means Technique and suppose there are k clusters. Then if for all the clusters , each point falls on an average in m canopies as before, there will be  mN/c exact distance comparisons in m different canopies since here only overlapping points are considered. Hence, for k clusters, the time complexity gets reduced to O(km(mN)/c) = O(Nk$m^{2}$/c) per iteration.

\vskip10pt Now, instead of canopies, if the Greedy Agglomerative clustering is performed on N data points, it requires O($N^{2}$) distance comparisons.
\vskip10pt If Expectation-Maximization or K-Means technique is performed on N data points and the number of clusters is k, then the computational complexity for those methods will be of O(Nk).
\vskip10pt Now m is the average number of canopies to which every data point belongs to.Hence m $\geq$ 1, and if there is no overlap, m=1. In addition, if the pair ($T_{1},T_{2}$) is chosen properly, then m is the smallest as compared to the number of observations. Hence, m $\ll$ N (or n). Since the number of observations is large and the number of overlaps kept as small as possible, then m $\sim$ 1. Then O($m^{2}N^{2}$/c) $\sim$ O($N^{2}$/c) which makes obvious that the more canopies are found, the more interesting the method is.
\vskip10pt It may also be noted that if the number of canopies are i.e., c is quite large and if the distance thresholds are selected properly, the number of overlapping observations will be negligible compared to the total number of observations. Hence, m$\ll$c, so that the fraction m$^{2}/c <$ 1.\\
It will imply that O($m^{2}$$ N^{2}$/c) $< $O($N^{2}$) and O(Nk$m^{2}$/c) $<$O(Nk).
\vskip10pt Thus, comparing all the methods discussed above, it can be said that the canopy technique reduces the complexity as a whole.

\vskip10pt Hence it may be concluded that canopy technique has the property that all points in any true cluster must fall in the same canopy
ensuring that no accuracy is lost by restricting comparisons of items to those in the same canopy.

\section{Spectra Data}

The spectra of 702248 galaxies and quasars with resdshift smaller
than 0.25 were retrieved from the Sloan Digital Sky Survey (SDSS)
database, release 7 (http://www.sdss.org/dr7/). Raw spectra have
3850 points in wavelength range imposed by the instrument,
$\lambda=3800$ and $\lambda=9250$ Angstrom. The spacing is
uniform in resolution ($\delta\lambda/\lambda)=1/4342$). However,
because of the redshifting of the spectra due to the expansion of
the Universe, the farthest objects of the sample (redshift of
0.25) have no data above the restframe wavelengths of
$9250/1.25$ =7400 Angstrom. The useful range of wavelengths after
correction for redshift is thus between 3806 and 7371 Angstrom.

\begin{figure}
     \includegraphics[width=0.5\textwidth]{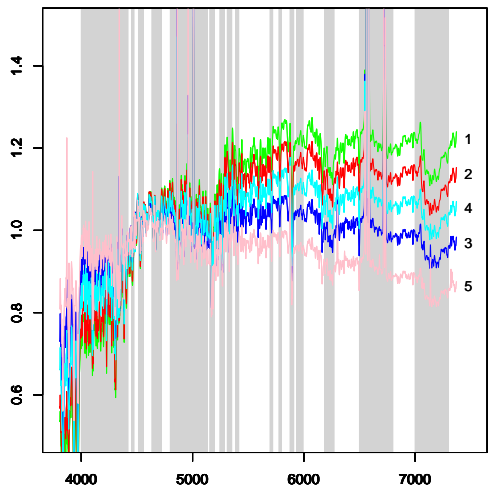}
\caption{Mean spectra for each cluster. The vertical gray stripes are the bands from which the 1540 parameters (spectrum wavelengths) were chosen.}
           \label{fig:groupspectra}
\end{figure}

To preserve the shape of spectral lines, the Shannon criteria has
been used while correcting for the redshift. For this purpose, we
doubled the sampling of the spectra beforehand, so that
 5740 points are obtained within the useful wavelength range after
redshift correction. Even for a large dimension this number is
quite large for Principal Component Analysis(PCA). For this we
then halved the sampling again, and selected the same wavelength
bands as in Sanchez-Almedia et al (2010; readers may see their
paper for the details and Fig.$~\ref{fig:groupspectra} $ for a
visual representation). These bands supposedly contain most of
the physics of galaxies, and reduce the number of points for each
spectra to 1539.

 Variables in the spectra cannot be standardized because they are of the same unit and are somehow related to each other.  We thus normalized the
 spectra globally with the flux average between $\lambda$ =4300 Angstrom et $\lambda$ =5000 Angstrom. We included the normalization factor as an additional parameter
  since the average level of each spectra reflects more or less the mass of the galaxy, which is an important physical property. This normalization
  factor of the global spectra is the only variable to be standardized in our analysis. The remaining question is to centralize or not the spectra
  (represented by the other 1539 variables). We believe that there is no absolute answer, and we considered both cases in our study. It should be
  remarked that a mean spectrum have no particular physical meaning, and like the spectrum of each galaxy which is a mixture of spectra coming from
  many stars and gaseous clouds, it merely reveals the average prominence of some features (stellar populations and atomic/molecular lines) in the
objects under study. The sample finally consists in 702248 spectra
with 1540 points (variables) for each.

\section{Data Analysis}
\subsection{KMeans Analysis}Principal Component Analysis(PCA) showed that the ¯rst four principal component account
for 85 percent of the variability and hence only the first four principal component are retained. Thus the data set is reduced to a matrix consisting
of 702248 rows and four columns
By performing a direct k-means analysis on the entire dataset (with respect to PC1,PC2,PC3
and PC4),three clusters are found. The cluster sizes are 2,513468 and 188778 respectively.\\


\vspace{0.1in} From table1 we see that the first cluster size is negligible, and the other two have significantly different mean spectra,
although one of the group contains 73 percent of the full
sample. However, the dispersion is large.
\vskip10pt Next another KMeans analysis has been performed on the entire data set with respect
to PC2,PC3 and PC4, i.e., by excluding PC1 which results in a more balanced cluster sizes i.e., 113888, 391589 and 196771 respectively.\\

\subsection{Choice of the cheap distance}
\vskip10pt Principal Component analysis showed that the first four
principal component account for 85 percent of the variability and
hence only the first four principal component are retained. Thus
the data set reduced to a matrix consisting of 702248 rows and
four columns representing the first four Principal Components of
the data sets respectively.

\begin{figure}
      \includegraphics[width=0.5\textwidth]{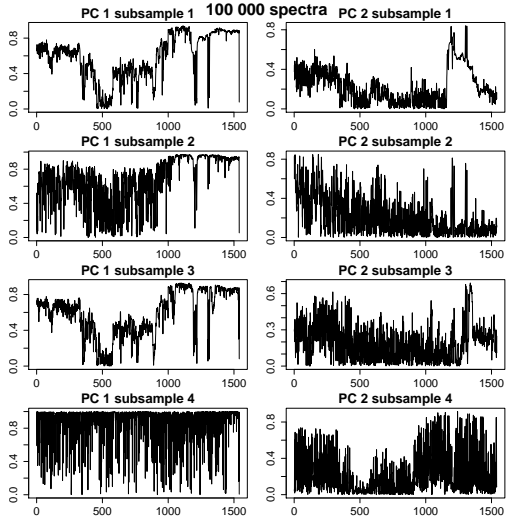}
  \caption{Eigenvectors (loadings) for the first and the second principal components for four mutually exclusive subsamples of 100~000 spectra.}
              \label{fig:loading1lacs}
\end{figure}

\begin{figure}
      \includegraphics[width=0.5\textwidth]{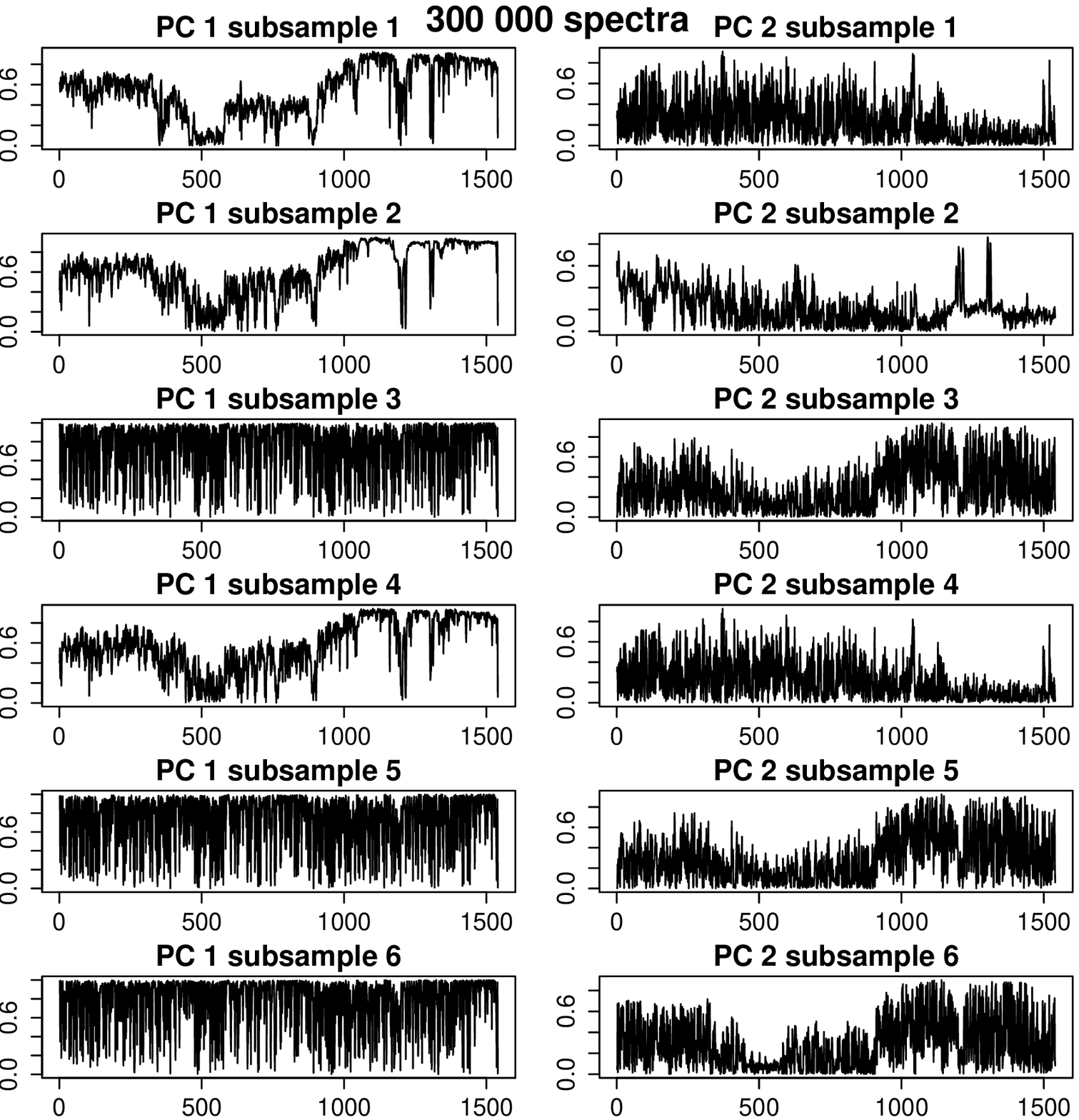}
  \caption{Eigenvectors (loadings) for the first and the second principal components for random subsamples of 300~000 spectra.}
              \label{fig:loading3.acs}
\end{figure}

\begin{figure}
      \includegraphics[width=0.5\textwidth]{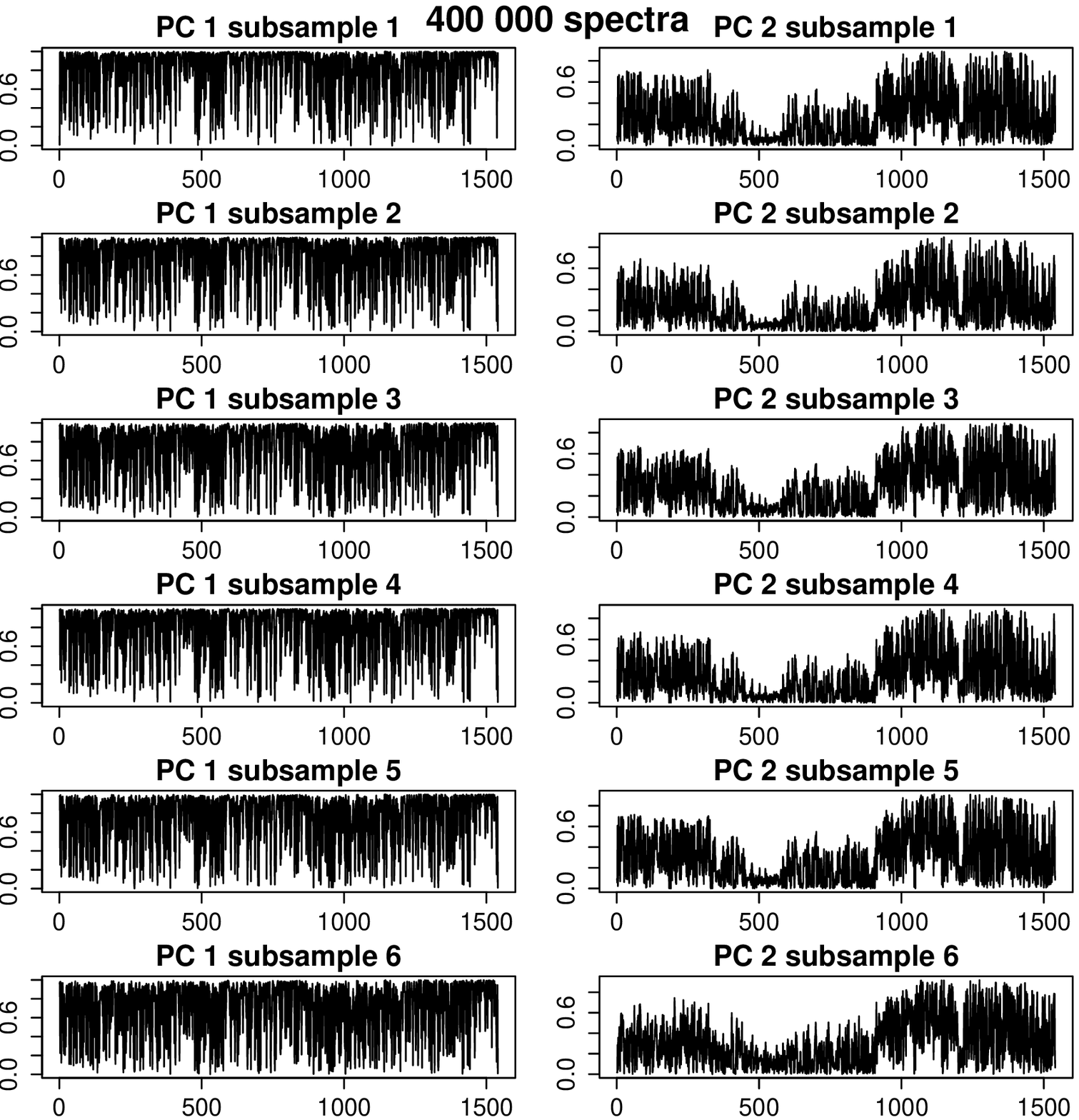}
   \caption{Eigenvectors (loadings) for the first and the second principal components for random subsamples of 400~000 spectra.}
              \label{fig:loading4lacs}
\end{figure}

\begin{figure}
      \includegraphics[width=0.5\textwidth]{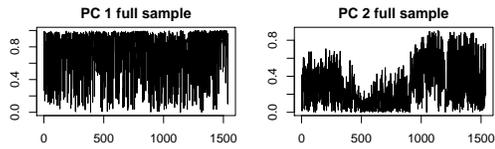}
   \caption{Eigenvectors (loadings) for the first and the second principal components for the full sample of 702248 spectra.}
              \label{fig:loading7lacs}
\end{figure}

One obvious choice for the cheap distance is the first principal component. It is the direction of maximum variance, even though not necessarily
the direction of highest clustering. However, by using subsamples, we have found that i) the loadings of PC1 and PC2 depend on the sample up to
a given size, ii) the first PC looses any discriminant role from 400000 objects upwards because nearly each variable reaches the maximum variance,
and iii) the second PC above 400000 objects resembles the first PC for some smaller samples
(Fig $~\ref{fig:loading1lacs}$,$~\ref{fig:loading3.acs}$,$~\ref{fig:loading4lacs}$,$~\ref{fig:loading7lacs}$). This confirms that,
in the present sample, the first PC indicates the direction of maximum variance, more and more clearly when the number of observations increases,
 but no clustering is present in this direction.From previous works on spectra data it has been found that mean subtraction ( "mean centering")
 is necessary for performing PCA to ensure that the first principal component describes the direction of maximum variance. If mean subtraction is
 not performed, the first principal component might instead correspond more or less to the mean of the data(Miranda  et al.\cite{Miranda} ). A mean of zero
 is needed for finding a basis that minimizes the mean square error of the approximation of the data.  Unsurprisingly, a canopy technique using
 the first PC as the cheap distance did not provide a satisfactory result. Four groups were found, but the mean spectra of each group were nearly
 identical, with large standard deviations.

Since the loadings of the second PC without centralizing the spectra is identical to the first PC with centralizing them, we performed a canopy
technique with the second PC as the cheap distance. For comparison, we also performed a k-means with the first PC and a k-means without the first PC, as discussed in the earlier section.

\subsection{Selection of distance thresholds $T_{1}$ and $T_{2}$}

After computing the cheap distance measures of all pairs of points
between the 2nd principal component column, it is found that the
mean of the distance column is 66.242.
\vskip10pt Hence the trial
pairs of values for $T_{1}$ and $T_{2}$ are taken as (65,58),
(72,63), (68,62) and (70,65). Then for each set of trial the
canopies (with overlapping observations) are formed and the proportion of correct observations,
precision and recall values are calculated. The results are given
in Tables 1,2,3 and 4.\\

From Table 1 we get, proportion correct = 0.767,total precision = 1.861,total recall = 2.144, 
 difference between total recall and total precision = 0.283.\\
From Table 2 we have, proportion correct = 0.821,total precision = 2.202,total recall = 2.429, 
difference between total recall and total precision = 0.227.

\vspace{0.1in}From Table 3, we get ,proportion correct = 0.888,total precision = 2.674, total recall =  2.651 
 , difference between total recall and total precision = 0.048.
  From table 4 we have, proportion correct = 0.935,total precision = 2.641, total recall =  2.815 
 , difference between total recall and total precision = 0.174.

\vskip10pt From the Tables 1,2,3 and 4 it is observed that the
proportion of correct grouping of observations is highest for the
pair (70, 65) i.e., 0.935.
 Also, the difference(0.174) between the total precision (2.815) and the total recall (2.815) is quite small. Although for the pair (68,62), the difference
 between the total recall and total precision is smallest (0.048), it can be seen that the proportion correct (0.888) is lesser than the previous choice.
 Also, with the choice (70,65), the number of overlaps reduces in a significant amount than the other choices considered.
Thus keeping in view all these points, the distance thresholds are
fixed at $T_{1}$ = 70 and $T_{2}$ = 65.

\subsection{ Subdivision of major canopies into further clusters}

With the above choices of the distance thresholds, 3 major canopies are formed having larger number of observations. The other canopies contains
 negligible observations and that can be ignored.


\vskip10pt Now, as it is seen from the above table that the total observation becomes 721209, meaning that there are some duplicate observations or
overlap. Now, to see whether elimination of overlaps increases the proportion correct or not, duplicate/repeated observations are ignored and only
mutually exclusive observations are considered. Then the sizes of the canopies become 453298, 202065 and 43533 respectively.The summary of
classification for these canopies is given in Table 5 which shows proportion correct = 0.955,total precision = 2.723, total recall =  2.871 
, difference between total recall and total precision = 0.148.\\
From Table 5, it may be seen that, after eliminating the
duplicate observations, the proportion of correct classification
increases. Also the precision and recall values are closer to
each other. Now, one point is, because of this elimination the
total number of observations becomes 698896, whereas, the actual
number of observations were 702248. But, since the proportion of
missing observations is insignificant,it can be
 ignored.


\vskip10pt In the next step, canopy 1 or C1 is divided into 2 clusters by K-means technique and the other canopies are kept as it is. Let the
clusters be named as K1, K2, K3, K4 .The summary of classification together with precision and recall values are shown in Table 6.

\vskip10pt Also, the subdivision of the 1st canopy into 2 clusters reduces the proportion correct. Thus, the next subdivision is done for the 2nd canopy, other canopies being kept as it is.
\\Let the clusters be denoted by G1, G2, G3, G4. The summary of classification and precision-recall values are shown in table 11.

\vspace{0.1in} Table 6 shows proportion correct = 0.917, total precision = 3.586 , total recall= 3.711 
,difference between total recall and total precision = 0.125

\vspace{0.1in} From Table 7, we get proportion correct = 0.923, total precision = 3.560, total recall = 3.638  
,difference between total recall and total precision = 0.078

\vskip10pt Here also the proportion of correct classification reduces from the proportion obtained without subdiving the canopies (refer to Table 5).
Also the difference between the precision and recall has been increased. Next both the subdivisions of canopy 1 and canopy 2 are taken together and canopy 3 is kept as is it.
Hence,now there are 5 clusters. The summary of classification for these clusters are given Table 8.

\vspace{0.1in} Table 8 gives proportion correct = 0.940, total precision = 4.685,  total recall = 4.743  
,difference between total recall and total precision = 0.058. From
table 12 it can be shown that the proportion of correct
classification is 0.94 whereas the proportion of correct
classification for 3 canopies (without subdivision) was 0.955. So,
the proportion is not reducing in a significant amount. But, on
the other end, the difference between total recall and total
precision is 0.058, whereas for 3 canopies the difference between
total recall and total precision was 0.148. So, this subdivision
is more appropriate in that case. \vskip10pt Finally the 5
clusters along with their sizes, WSS and descriptive measures are
given in Table 9.

\vskip10pt Table 9 shows that the within cluster sum of squares reduces in a significant amount by subdividing the 1st and 2nd canopy. Also, the proportion correct
is quite high (0.94) and the total recall and the total precision are closest to each other. Considering all these facts, the number of clusters
is finally taken as k = 5.

\section{Interpretation of the Clusters}

The mean spectra for clusters 1 to 5 appear to be ordered according to the ratio between the blue and red parts, cluster 1 being the reddest and cluster 5
the bluest (Fig.$~\ref{fig:groupspectra}$). Hence, the stellar population on average is younger from cluster 1 to cluster 5.

\begin{figure}
      \includegraphics[width=0.5\textwidth]{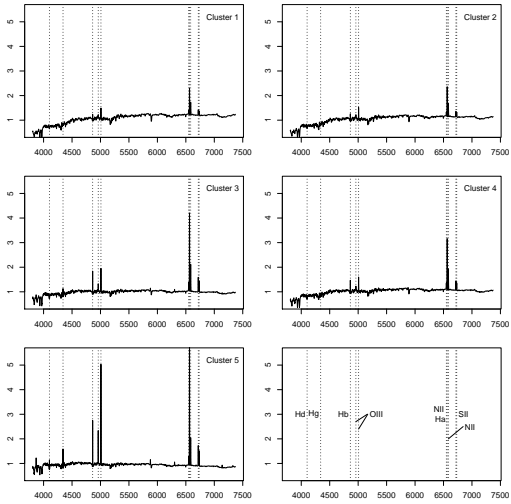}
   \caption{Mean spectra for each cluster with identification of emission lines.}
              \label{fig:groupspectralines1}
\end{figure}

This is corroborated by the stronger emission lines in cluster 5 (Fig.$~\ref{fig:groupspectralines1}$) which are characteristic of young stellar
populations. However, it is interesting to note that cluster 3 has stronger emission lines than cluster 4 which is bluer. This could be probably
interpreted as a different mixture of stellar populations and a different proportion of ongoing star formation.

\begin{figure}
      \includegraphics[width=0.5\textwidth]{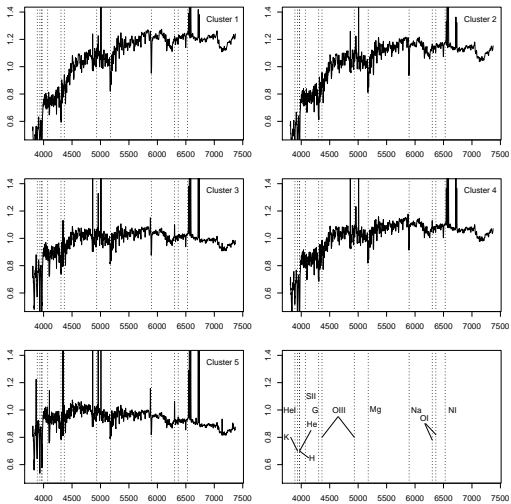}
   \caption{Mean spectra for each cluster with identification of absorption lines.}
              \label{fig:groupspectralines2}
\end{figure}

Stronger absorption features are typical of older stellar populations, and the mean spectra are fully consistent with this picture
(Fig.$~\ref{fig:groupspectralines2}$). cluster 1 and cluster 2 have the oldest stellar populations, and differ from the global slope, cluster 2 mean
spectrum being slightly flatter,  and from a Na absorption feature that is slightly deeper for cluster 2. In addition, the absorption features
are clearly deeper in cluster 4 than in cluster 3.

It is not in the scope of the present paper to detail the
astrophysical interpretation, but it is already clear that a
combination of canopy and k-means techniques applied to more than
700~000 raw spectra were able to distinguish five groups which
have clear specific and consistent physical properties. Five
classes is probably not enough to fully describe the huge
diversity of galaxies in the Universe. However, there are at least
three reasons why this number is reasonable. Firstly, the raw
spectra as we used them provide information only on star colours
and gaseous tracers. Our study ignores the dynamical properties of
the galaxy constituents (given by the width of the emission or
absorption lines), as well as more detailed properties of physical
conditions, such as the temperature and radiative environment of
the gas clouds, that are given by line ratios. These measures
require model-fitting of specific regions of the spectra.
Secondly, the spectrum of a galaxy comes from a mixture of gas
clouds and of a huge number of stars from several
distinct populations with distinct properties. Such global spectra
have a higher probability of looking more or less similar.
Thirdly, chemical and physical processes that explain galaxy
diversity are continuous processes. They mainly yield a continuum
of variance for each variable, hence the large number of spectra
of our sample is certainly rather homogeneous in the parameter
space, making clustering quite difficult.

\section{Discussion}

Sanchez Almeida (2010)\cite{Sanchez Almeida} used a k-means
analysis and found several classifications. They note that their
k-means result depends very much on the initial seeds. Using some
criteria, they end up with four classifications, and chose one
arbitrarily. This partitioning has 28 clusters and subsequently
they gave some astrophysical meaning to them.

Our study is clearly in disagreement with their work.Our analysis
is free from seed dependence problem as we have used the method
proposed by Milligan, (1980)\cite{Milligan}. Being much more
robust, our results never find several possible partitioning and
in all cases only a very small number of clusters is found if any
at all, even when using a direct k-means computation as they did.
We explain this disagreement(very small number of clusters:5
instead of 28) in two ways. Firstly, we are probably facing a
curse of dimensionality with this large sample with a large number
of variables. This problem is strengthened
 by the homogeneity of the observations in the parameter space. Spectra of galaxies result from continuous physical processes and increasing the number of observations make the spectra to fill the entire space of possible occurrences. Secondly, each spectrum is a mixture of an assembly of billions of
  stars and gas  clouds. In other words, the clustering of extragalactic spectra is very probably not prominent at all, or it is blurred by the too many
  variables present in spectra.

The PCA have helped us to select the most discriminant variables, but even so only three groups have been identified. In addition, and may be more
importantly, we have shown how much the principal components depend on the sample and on the sample size. One should not forget that PCA works well
when there is much more observations than variables. Hence building general classifications, or identifying physical groups in the eigenvector space,
can be misleading, unless a kind of cross-validation is made as we have done in this paper.

\section{Conclusion}

From the present work we may draw the following conclusions:
\begin{enumerate}
 \item The most important finding is that for a large number of
 high dimensional data points the proposed method is a very good
 technique for clustering.
 \item Even PCA performs differently with large data and one
 should take care of those limitations.
 \item In astronomy, analysis of spectra data is very common and
 for proper statistical analysis one should take care of different
 Astronomical as well as statistical properties.
 \item For canopy technique the choice of the "cheap distance
 measure" is an important issue and proper choice will lead to a
 good answer.
\end{enumerate}


\section{Acknowledgement}The first author Tuli De acknowledges her work
to a research fellowship provided by The University Grants
Commission , sanction No. $UGC/157/JrFellow(sc)$ dated $31/03/2010$.
The authors are also grateful to Mr. Sourov Ghosh, Advisory
Consultant, IBM India and Mr. Bharat Warule,Analytical Developer
and Implementer, Cypress Analytica ,Pune for their help in some of
the computations.



\newpage


\begin{table}
\begin{tiny}
 \begin{center}
\textbf{Table1:}\textbf{Summary of classification of canopies with precision and recall for $T_{1}$ =65 and $T_{2}$= 58::}\\
 \end{center}
\begin{tabular}{|p{1in}|p{.4in}|p{.4in}|p{.4in}|p{.7in}|p{.7in}|p{.7in}|}
\hline
{\textbf{Put into Groups}} & \multicolumn{6}{c|}{\textbf{True Groups}}  \\
\cline{2-7}
&   1 &   2 &3 & Total N & N correct & Precision\\
\hline
\textbf{ 1}  &  478326   &    1962 &   2681 &   482969 &   478326  &    0.99 \\
\hline
\textbf{ 2}  &  36949  &  87278   &  6963     &  131190   &   87278   &  0.665 \\
\hline
\textbf{ 3}  &   58056    &   76668 &   35044  &  169768  &   35044 &   0.206 \\
\hline
\textbf{Total N}  &   573331  &  165908 &   44688 &  -  &  - & -  \\
\hline
\textbf{N Correct} &    478326  &  87278  &  35044 & - & -& - \\
\hline
\textbf{Recall} &   0.834  &  0.526  &  0.784 & - & -& - \\
\hline
\end{tabular}
\end{tiny}
\end{table}

\begin{table}
\begin{tiny}
\begin{center}
\textbf{Table2:}\textbf{Summary of classification into canopies with precision and recall for $T_{1}$ =72 and $T_{2}$= 63:}\\
 \end{center}
\begin{tabular}{|p{1in}|p{.4in}|p{.4in}|p{.4in}|p{.7in}|p{.7in}|p{.7in}|}
\hline
{\textbf{Put into Groups}} & \multicolumn{6}{c|}{\textbf{True Groups}}  \\
\cline{2-7}
&   1 &   2 &3 & Total N & N correct & Precision\\
\hline
\textbf{ 1}  &  403321   &    32861 &   3524 &   439706 &   403321  &    0.917 \\
\hline
\textbf{ 2}  &  68247  &  190674   &  7058  &  265979   &   190674   &  0.717 \\
\hline
\textbf{ 3}  &   17958 &  12256 &  39770  &  69984  &  39770 &   0.568 \\
\hline
\textbf{Total N}  &   485926  &  235791 &   50352 &  -  &  - & -  \\
\hline
\textbf{N Correct} &  403321  &  190674  &  39770 & - & -& - \\
\hline
\textbf{Recall} &   0.830  &  0.809  &  0.790 & - & -& - \\
\hline
\end{tabular}
\end{tiny}
\end{table}

\begin{table}
\begin{tiny}
\vspace{0.1in}
\begin{center}
\textbf{Table3:}\textbf{Summary of classification of canopies with precision and recall for $T_{1}$ =68 and $T_{2}$= 62:}\\
\end{center}
\begin{tabular}{|p{1in}|p{.4in}|p{.4in}|p{.4in}|p{.7in}|p{.7in}|p{.7in}|}
\hline
{\textbf{Put into Groups}} & \multicolumn{6}{c|}{\textbf{True Groups}}  \\
\cline{2-7}
&   1 &   2 &3 & Total N & N correct & Precision\\
\hline
\textbf{ 1}  &  384047   &   33369 &   1893 &  419309 &  383047 &    0.913 \\
\hline
\textbf{ 2}  &  28616  &  297708   &  2873  &  249197   &   217708   &  0.874 \\
\hline
\textbf{ 3}  &   4135 &  9459 &  61845  &  75439  &  61845 &   0.820 \\
\hline
\textbf{Total N}  &   416798  &  260536 &   66611 &  -  &  - & -  \\
\hline
\textbf{N Correct} &  384047  &  207708  &  61845 & - & -& - \\
\hline
\textbf{Recall} &   0.921  &  0.836  &  0.928 & - & -& - \\
\hline
\end{tabular}
\end{tiny}
\end{table}

\begin{table}
\begin{tiny}
\begin{center}
\textbf{Table4:}\textbf{Summary of classification into canopies with precision and recall for $T_{1}$ =70 and $T_{2}$= 65:}\\
 \end{center}
\begin{tabular}{|p{1in}|p{.4in}|p{.4in}|p{.4in}|p{.7in}|p{.7in}|p{.7in}|}
\hline
{\textbf{Put into Groups}} & \multicolumn{6}{c|}{\textbf{True Groups}}  \\
\cline{2-7}
&   1 &   2 &3 & Total N & N correct & Precision\\
\hline
\textbf{ 1}  &  438421   &   15237 &   0 &  453658 &  438421 &    0.966 \\
\hline
\textbf{ 2}  &  14874  &  186535   &  336  &  201745   & 186535   &  0.925 \\
\hline
\textbf{ 3}  &   3 &  16454 &  49349  &  65806  &  49349 &   0.750 \\
\hline
\textbf{Total N}  &   453298  &  218226 &   49685 &  -  &  - & -  \\
\hline
\textbf{N Correct} &  438421  &  186535  &  49349 & - & -& - \\
\hline
\textbf{Recall} &   0.967  &  0.855 &  0.993 & - & -& - \\
\hline
\end{tabular}
\end{tiny}
\end{table}



\begin{table}
\begin{tiny}
\begin{center}
\textbf{Table5:}\textbf{Summary of classification into 3 canopies with precision and recall after eliminating overlaps:}\\
 \end{center}
\begin{tabular}{|p{1in}|p{.4in}|p{.4in}|p{.4in}|p{.7in}|p{.7in}|p{.7in}|}
\hline
{\textbf{Put into Groups}} & \multicolumn{6}{c|}{\textbf{True Groups}}  \\
\cline{2-7}
&   1 &   2 &3 & Total N & N correct & Precision\\
\hline
\textbf{ 1}  &  441348   &   8502 &   0 &  449850 &  441348 &    0.981 \\
\hline
\textbf{ 2}  &  11947  &  183094   &  387  &  195428   & 183094   &  0.937 \\
\hline
\textbf{ 3}  &   3 &  10469 &  43146  &  53618  &  43146 &   0.805 \\
\hline
\textbf{Total N}  &   453298  &  202065 &   43533 &  -  &  - & -  \\
\hline
\textbf{N Correct} &  441348  &  183094  &  43146 & - & -& - \\
\hline
\textbf{Recall} &   0.974  &  0.906 &  0.991 & - & -& - \\
\hline
\end{tabular}
\end{tiny}
\end{table}


\begin{table}
\begin{tiny}
\begin{center}
\textbf{Table6:}\textbf{Summary of result of subdivision of canopy 1 into two clusters with precision and recall }\\
\end{center}
\begin{tabular}{|p{1in}|p{.4in}|p{.4in}|p{.4in}|p{.5in}|p{.6in}|p{.7in}|p{.5in}|}
\hline
{\textbf{Put into Groups}} & \multicolumn{7}{c|}{\textbf{True Groups}}  \\
\cline{2-8}
&   1 &   2 &3 &4 & Total N & N correct & Precision\\
\hline
\textbf{ 1}  &    152039  &   7939 &    0  &      0  &      159978  &   152039  &  0.950\\
\hline
\textbf{ 2}  &     16559 &   274601  &    17879   &    0  &  309039 &   274601 &   0.888 \\
\hline
\textbf{ 3}  &    10   &    2150  &  170503  &    1  &   172663  &   170503 &  0.987 \\
\hline
\textbf{ 4}  &    0  &   0  &   13683  &  43532  &    57215  &   43532 &   0.761 \\
\hline
\textbf{Total N}      &   168608 &  284690  &  202065  &  43533 &-& - & -\\
\hline
\textbf{N Correct}   &  152039 &  274601 &   170503  &  43532 & - & -& - \\
\hline
\textbf{Recall}   &   0.902  &  0.965  &  0.844  &  1.000  &  - & -& - \\
\hline
\end{tabular}
\end{tiny}
\end{table}

\begin{table}
\begin{tiny}
\begin{center}
\textbf{Table7:}\textbf{Summary of result of subdivision of canopy 2 into two clusters with Precision and recall:}\\
\end{center}
\begin{tabular}{|p{1in}|p{.4in}|p{.4in}|p{.4in}|p{.5in}|p{.6in}|p{.7in}|p{.5in}|}
\hline
{\textbf{Put into Groups}} & \multicolumn{7}{c|}{\textbf{True Groups}}  \\
\cline{2-8}
&   1 &   2 &3 &4& Total N & N correct & Precision\\
\hline
\textbf{ 1}  &    418073   &    12  &   1264 &   0    &   419349   &   418073   &    0.997\\
\hline
\textbf{ 2}  &    13  &  81825  &  5320  &    7650   &  94808   &   81825   &  0.863  \\
\hline
\textbf{ 3}  &   35210 &  2804 &  109282  &   0   &  147296    &  109282   &   0.742 \\
\hline
\textbf{ 4}  &   2  &    1558  &   0  &  35883  &    37443 &   35883  &    0.958 \\
\hline
\textbf{Total N}      &   453298  & 86199  &  115866  & 43533   & -& - & -\\
\hline
\textbf{N Correct}      &  418073  &  81825  &  109282 &  35883   & - & -& - \\
\hline
\textbf{Recall}      &  0.922 &   0.949  &  0.943 &   0.824 & - & -& - \\
\hline
\end{tabular}
\end{tiny}
\end{table}

\begin{table}
\begin{tiny}
\begin{center}
\textbf{Table8:}\textbf{Summary of result of subdivision of canopy 1 and canopy 2\\
 into 2 clusters each with precision and recall:}\\
\end{center}
\begin{tabular}{|p{1in}|p{.4in}|p{.4in}|p{.4in}|p{.4in}|p{.5in}|p{.6in}|p{.7in}|p{.5in}|}
\hline
{\textbf{Put into Groups}} & \multicolumn{8}{c|}{\textbf{True Groups}}  \\
\cline{2-9}
&   1 &   2 &3 &4& 5 & Total N & N correct & Precision\\
\hline
\textbf{ 1}  &    160511   &    5027  &   0 &   0    & 0&   165538  &   160511  & 0.970\\
\hline
\textbf{ 2}  &    8083  &  264248  &  36  &  2679   &  0 & 275046   &  264248  &  0.961  \\
\hline
\textbf{ 3}  &   2 &  0 & 83005  &   4339   & 3427 &  90773   &  83005   &  0.914 \\
\hline
\textbf{ 4}  &   12  &    15415  &   2394  &  108848  &   0 &  126669 &   108848  &  0.859 \\
\hline
\textbf{ 5}  &   0  &    0  &   764  &  0  &    40106 &  40870 & 40106  &  0.981 \\
\hline
\textbf{Total N}      &   168608  & 284690  &  86199  & 115866 & 43533   & -& - & -\\
\hline
\textbf{N Correct}      &  160511  &  264248  &  83005 & 108848 & 40106  & - & -& - \\
\hline
\textbf{Recall}      &  0.952 &   0.968  &  0.963 &   0.939 & 0.921 & - & -& - \\
\hline
\end{tabular}
\end{tiny}
\end{table}

\begin{table}
\begin{tiny}
\begin{center}
\textbf{Table9:}\textbf{Canopies with sizes,wss and descriptive measures:}\\
\end{center}
\begin{tabular}{|c|c|c|c|c|c|c|c|c|c|c|}
\hline
$Canopy$ & $Sizes$ & $wss$ & \multicolumn{2}{c|}{$PC1$} & \multicolumn{2}{c|}{$PC2$}& \multicolumn{2}{c|}{$PC3$}& \multicolumn{2}{c|}{$PC4$}\\
 \cline{4-11}
 &  &  & $Mean$ & $S.E. Mean$ & $Mean$ & $S.E. Mean$ & $Mean$ & $S.E. Mean$ & $Mean$ & $S.E. Mean$ \\
 \hline
\textbf{ C1} & 168608 & 103153104.005 & 235.21 &  0.03 & 296.41 & 0.05 & 17.650 & 0.006 & 102.47 & 0.02\\
\hline
\textbf{ C2} & 284690 & 112477167.915 & 205.41 &  0.02 & 258.31 & 0.02 & 16.976 &  0.005 & 98.749 &  0.015\\
\hline
\textbf{ C3} & 86199  & 64095947.252  & 115.53 & 0.07 & 160.20 &  0.05 & 21.417 &  0.011 & 107.33 & 0.02 \\
\hline
\textbf{ C4} & 115866 & 50857574.152  & 162.80 & 0.04 & 208.51 &  0.04 & 19.412 &  0.008 & 104.75 & 0.02 \\
\hline
\textbf{ C5} & 43533  & 468132949.793 & 30.06  & 0.48 & 102.84 &  0.11 & 25.064 & 0.038 & 109.43 & 0.05 \\
\hline
\end{tabular}
\end{tiny}
\end{table}


\end{document}